\documentclass[twocolumn,showpacs,preprintnumbers,amsmath,amssymb,superscriptaddress]{revtex4}
\usepackage{graphicx}%
\usepackage{dcolumn}%
\usepackage{bm}%
\usepackage{color}
\usepackage{subfigure}
\usepackage[utf8]{inputenc}
\usepackage{layouts}
\usepackage{todonotes}
\usepackage{hyperref}
\hypersetup{colorlinks=true,citecolor=blue}

\begin{document}

\title{Effective low-dimensional dynamics of a mean-field coupled network of slow-fast spiking lasers.}

\author{A.~Dolcemascolo}
\author{A.~Miazek}
\affiliation{Universit\'e C\^ote d'Azur, CNRS, INPHYNI, 1361 route des lucioles 06560 Valbonne, France}
\author{R.~Veltz}
\affiliation{Inria Sophia Antipolis, MathNeuro Team, 2004 route des Lucioles - BP93, 06902 Sophia Antipolis, France}
\author{F.~Marino}
\affiliation{CNR-Istituto Nazionale di Ottica and INFN, Sez. di Firenze, Via Sansone 1, I-50019 Sesto Fiorentino (FI), Italy} 
\author{S. Barland}
\affiliation{Universit\'e C\^ote d'Azur, CNRS, INPHYNI, 1361 route des lucioles 06560 Valbonne, France}

\date{\today}
\newcommand{\romain}[2][noinline]{\todo[#1, color=blue!20!white]{\small \texttt{Romain}: #2}}
\newcommand{\stephane}[2][noinline]{\todo[#1, color=blue!5!white]{\small \texttt{Stephane}: #2}}

\begin{abstract}
Low dimensional dynamics of large networks is the focus of many theoretical works, but controlled laboratory experiments are comparatively very few. Here, we discuss experimental observations on a mean-field coupled network of hundreds of semiconductor lasers, which collectively display effectively low-dimensional mixed mode oscillations and chaotic spiking typical of slow-fast systems.  We demonstrate that such a reduced dimensionality originates from the slow-fast nature of the system and of the existence of a critical manifold of the network where most of the dynamics takes place. Experimental measurement of the bifurcation parameter for different network sizes corroborate the theory.
\end{abstract}

\pacs{Valid PACS appear here}
                             
\maketitle

\newcommand{\fig}[1]{Fig.~\ref{#1}}

\makeatletter
\@input{modtime}
\makeatother

The collective dynamics of large ensembles of coupled systems is a far reaching research topic and striking natural examples of reduced dynamics dimensionality in large networks abound, like fireflies or applause synchronization \cite{neda2000physics}. One paradigmatic  example is the synchronization of globally coupled phase oscillators as observed in the Kuramoto model \cite{kuramoto2012chemical}, whose relative simplicity has allowed tremendous progress (see \textit{e.g.} \cite{strogatz2000kuramoto}). Beyond this idealistic case, a particularly relevant situation is that of spiking nodes such as neurons, whose synchronization may play a key role in epilepsy \cite{jiruska2013synchronization}. Thus, many studies focus on the reduced dimensionality of the dynamics of networks of neuron models, see \textit{e.g.} \cite{mirollo1990synchronization,watanabe1994constants,zillmer2006desynchronization,olmi2014hysteretic,kotani2014population,montbrio2015macroscopic,pazo2016quasiperiodic}, often enabled by the so-called Ott-Antonsen ansatz \cite{ott2008low,ott2009long}. In contrast to this rich theoretical literature, experimental observations are scarce. Here, we study the dynamics of a mean-field coupled network of chaotically spiking, dynamically coupled semiconductor lasers. We observe experimentally mixed mode oscillations and chaotic spiking in the mean field, which result from partial synchronization along the slow manifold of the network even in absence of synchronization of the fast dynamics of the nodes.

The analysis of optical model systems is often  useful in nonlinear science, in particular about the synchronization of oscillators as shown in lasers in \cite{nixon2011synchronized,nixon2012controlling}. With respect to neurosciences, optical analogues of neurons abound (recent references include  \cite{selmi2014relative,hurtado2015controllable,sorrentino2015effects,mesaritakis2016artificial,prucnal2016recent,dolcemascolo2018resonator}) but only very few nodes have been experimentally coupled: self-coupling with delay in \cite{garbin2015topological,romeira2016regenerative,terrien2018pulse}, and two nodes in \cite{yacomotti2002coupled,kelleher2010excitation,van2012cascadable,deng2017controlled}. In contrast, we study a large network of 451 elements. The coupling is dynamic, mimicking \textit{pulse-coupled} networks \cite{belykh2005synchronization}, and the topology can be experimentally tuned from one to all to fully connected. Each of the nodes is a three-dimensional slow-fast system producing relaxation- and mixed mode oscillations and chaotic spiking. 

Although the mean field cannot be described by an ordinary differential equation, we observe an effectively low dimensional dynamics of the network due to the slow-fast nature of the system. Most of the dynamics takes place close to a simple critical manifold whose stability can be computed analytically. The convergence of a bifurcation parameter towards a unique value is observed experimentally by increasing the network size in a quenched disorder configuration.

\begin{figure}[h!]
\includegraphics[width=0.47\textwidth]{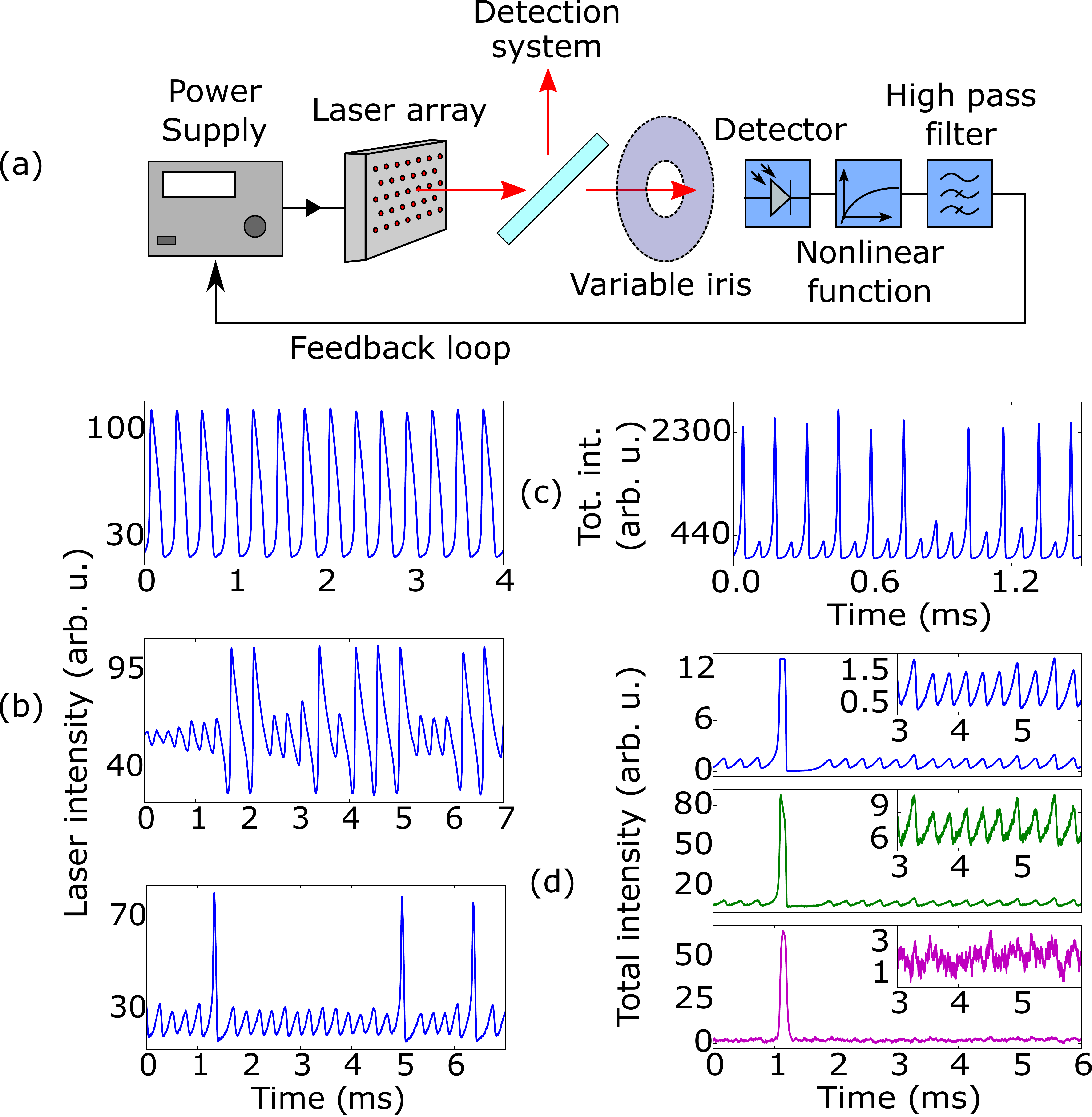}
	\caption{Experimental setup and typical dynamics. (a) The light emitted by an array of 451 lasers is converted to an electrical signal which, after filtering and nonlinear transformation, is reinjected into the single current source driving all lasers. (b) Single-laser time-traces showing: periodic oscillations, chaotic bursting and spiking (pumping currents 195.0~mA, 196.5~mA and 187.0~mA). (c) Mean-field time-trace of 451 lasers (182.1 mA), showing MMOs. (d) Top panel: total intensity of all 451 lasers, middle and bottom panels: intensity of two different lasers (pumping is 189.9~mA). 
	\label{fig:setup}
	}
\end{figure}

The experiment is shown on \fig{fig:setup}a). An array of 451 semiconductor lasers (Vertical Cavity Surface Emitting Lasers, VCSELs) is submitted to an AC-coupled nonlinear optoelectronic feedback. The dynamics of a single semiconductor laser can be described by two real coupled variables of widely differing time scales (light intensity, 10~ps, and semiconductor carrier population, 1~ns). As shown in \cite{al2009chaotic,marino2011mixed}, chaotic spiking can arise via an incomplete homoclinic snaking scenario when a laser is driven close to its first (transcritical) bifurcation point and when an electric signal proportional to the intensity of the emitted field is reinjected back into the pumping current after a saturable nonlinear transformation and high-pass filtering. This signal constitutes a third (still much slower, typically 1 ms) variable. 
Due to the large pitch between the lasers there is no nearest neighbor coupling and the wavelength distribution of the lasers spans over 2~nm, preventing coherent interactions. All the lasers are driven by a single power supply, whose current is distributed evenly between all lasers (of identical impedance). The threshold current distribution is symmetric with average value 183.3~mA and standard deviation 5.8~mA. The emitted light is collected by a short focal length lens which forms an image of the array after about 10~cm propagation. Slightly before the image plane, the beam is split in two, one for detection and the other for the opto-electronic reinjection. In this beam, at the image plane, a variable aperture iris is used to control the sub-population which drives the dynamics. The light emitted by this population is converted by a photodetector into a voltage which is logarithmically amplified, providing a saturable nonlinearity. The continuous component is actively filtered out and the resulting signal is sent as a control voltage into the laser power supply. The aperture of the iris controls the coupling, from one to all to globally coupled. The control parameters are the driving current and the amount of light sent to the detector (controlled via a neutral density filter).

When the iris is closed to select a single laser, this device's intensity drives the current applied to the whole population. The intensity of that particular laser can display complex dynamics as in \cite{al2009chaotic,marino2011mixed} including relaxation oscillations and chaotic bursting or spiking as shown in \fig{fig:setup}b). When the iris is completely open, the total intensity drives the power supply pumping the whole array, resulting in a mean-field coupled network of 451 nodes. Strikingly, the network can display periodic and chaotic mixed mode oscillations (MMOs) as shown in \fig{fig:setup}c). On \fig{fig:setup}d) we show synchronous measurements of the total intensity and of the intensity  emitted by two different lasers in the mean-field coupled configuration during chaotic spiking: both lasers spike when the network spikes, but only one laser (central trace) displays the sub-threshold oscillations observed at the network level (top trace), while the other laser remains quiet (at the detection noise level, bottom trace).

\begin{figure}[h]
\includegraphics[width=0.45\textwidth]{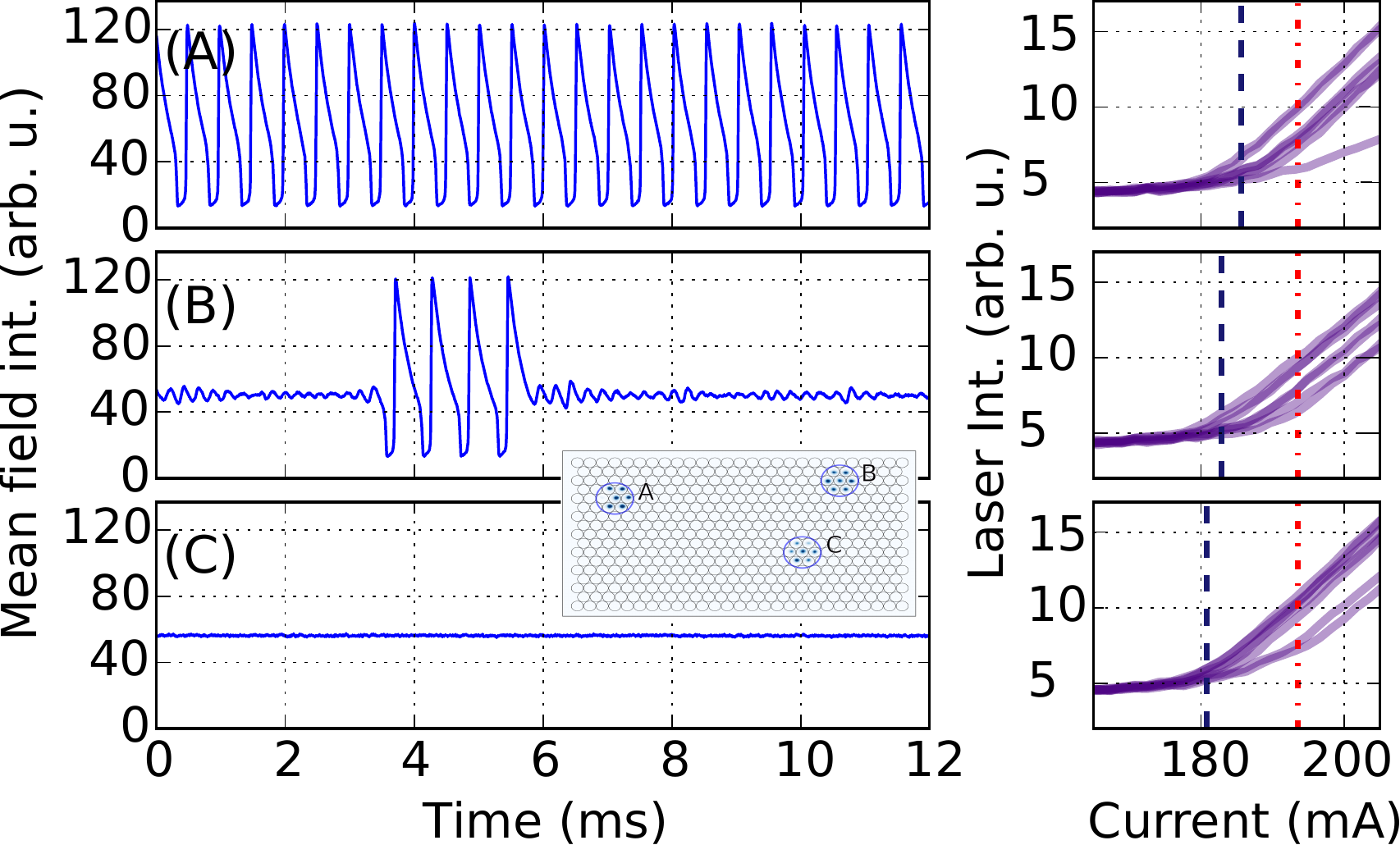}
	\caption{Dynamics of small networks. Different sub-populations (A, B and C) can be selected, showing different dynamics (left column). Each uncoupled population has a different laser threshold distribution (right column), the black dashed line shows the average threshold for each population (A:185.63~mA, B:182.88~mA, C:180.81~mA). The red dash-dotted line shows the constant current value used in all measurements (193.50~mA). 
	\label{fig:seven} 
	}
\end{figure}

Smaller networks can be studied by partly closing the iris and detecting the corresponding population. Different dynamics are observed depending on the sub-population (\fig{fig:seven}). All parameters are constant and the amount of light sent to the detector is scaled to maintain the coupling constant when changing population. Each sub-population consists of seven elements whose threshold current differs slightly from device to device. Network A shows relaxation oscillations, B shows chaotic bursting and C is stationary. In all cases the dynamics of the total intensity seems low-dimensional. Each population is characterized by its threshold current distribution. The existence of a well-identified low-dimensional dynamics in populations of identical size but with distinct average threshold suggests that this parameter controls the dynamics.

The emergence of chaotic MMOs and of an effective, low-dimensional mean field dynamics can be inferred from the following. We consider a population of $N$ semiconductor lasers globally coupled through a common AC-coupled optoelectronic feedback. Each laser is modelled by  standard single-mode rate equations describing the evolution of the optical intensity, carriers and feedback current. After proper scaling \cite{si}, the equations read:
\begin{eqnarray}\label{eq:model1}
\dot{x_i} & = & x_i(y_i-1) \;   \\
\dot{y_i} & = & \gamma(\delta_i-y_i+k(w+f(X)) - x_iy_i) \;  \label{eq:model2} \\
\dot{w} & = & -\epsilon(w + f(X)) \;  \label{eq:model3} 
\end{eqnarray}
where time has been normalized to the photon lifetime and $x_i, y_i$ are respectively the dimensionless photon and carrier density of the laser $i$ and $X = \frac{1}{N}\sum\limits_{i=1}^N x_i$ is the total intensity normalized to the number of elements. The global variable $w$ is the (scaled) high-pass filtered feedback current, which includes a saturable nonlinear function $f(X) = A\ln(1+\alpha X)$. The optical and electrical propagation delays are negligible.
All the lasers are considered identical except for the coherent emission threshold current that is included in the control parameter $\delta_i$ (proportional to the ratio between the common pump and the threshold current of each laser).

For $N=1$, there are two equilibria $(0,\delta_1,0)$, $(\delta_1-1,1,-f(\delta_1-1))$.
Since the normalized carrier rate $\gamma$ and AC feedback cutoff frequency $\epsilon$ are such that $\epsilon \ll \gamma \ll 1$, Eqs. (\ref{eq:model1}-\ref{eq:model3}) is a slow-fast system with three timescales. 
This model is strongly reminiscent of that of  \cite{al2009chaotic,al2010excitability,marino2011mixed}. Similarly, the slow dynamics take place near a one-dimensional manifold $\Sigma = \Sigma_x \cup \Sigma_y$, where the lower attractive branch $\Sigma_x$ is given by the zero-intensity solution $\Sigma_x=\{x = 0, y_w = \delta_1 + k w, w \}$ while the middle repulsive and upper attracting branch, $\Sigma_y = \{x_w, y = 1, w \}$, is implicitly defined by the equation $\delta_1 - 1 + k w + k f(x_w) - x_w = 0$. Since two branches rapidly attract all neighboring trajectories while the middle branch repels them, canard and relaxation cycles arise. 
These features are common in planar slow-fast systems but here a third intermediate time-scale, $1/\gamma$, induces more complex scenarios. First, the fixed points of the 2D fast subsystem (\ref{eq:model1}-\ref{eq:model2}) laying on the upper attractive critical branch consist of stable foci. Therefore the trajectories near these branches are shrinking helicoids, in contrast with the monotonic decay of the planar case. Second, a regime of regular or chaotic MMOs takes place, where canard orbits are separated by small-amplitude, quasi-harmonic oscillations surrounding the steady state of the system. When laying on the middle repelling branch, such equilibrium is a saddle focus and trajectories can rotate several times around it before switching to the other stable branch of the manifold. The number of these rotations, as well as the periodic or erratic nature of MMOs \cite{desroches2012mixed}, are determined by the rates at which both $y_1$ and $w$ vary in the vicinity of the saddle-focus. This is related to the values of $\gamma$ and $\epsilon$, but also critically depends on the bifurcation parameter $\delta_1$.

When $N>1$, (\ref{eq:model1}-\ref{eq:model3}) describe a network of  $N$ such elements, globally coupled through their slowest variable $w$.
It admits $2^N$ equilibria which can be computed by splitting the population using the set $I$ of $N_+$ switched ON lasers with $(x_i,y_i)=(\delta_i-1,1),\ i\in I$ and the others $(x_i,y_i)=(0,\delta_i),\ i\notin I$ with $w=-f(\frac1N\sum\limits_{i\in I}\delta_i-\frac{N_+}{N})$. The stability of these equilibria can be computed as function of $\delta_i$ defining $\Delta = \frac{1}{N}\sum\limits_{i=1}^N \delta_i$ and assuming $\delta_i = \Delta+z\eta_i, z\ll1$, \textit{i.e.} that the lasers are similar enough to each other \cite{si}. At zero order in $z$, only the proportion $\frac{N_+}{N}$ of switched ON lasers affects the stability of the network which is otherwise determined by~$\Delta$.

\begin{figure}[t]

	\includegraphics[width=0.45\textwidth]{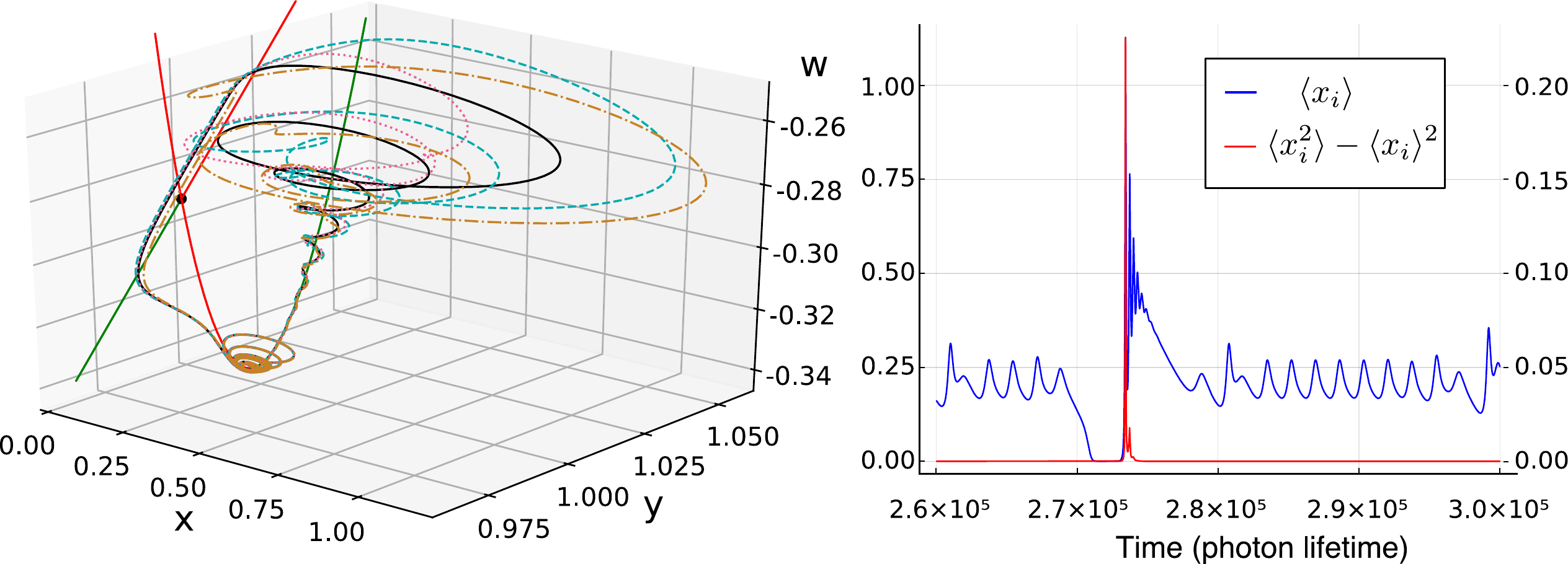}%
	\caption{Numerical simulations of $10^4$ coupled lasers with optoelectronic feedback modelled by \eqref{eq:model1}. Only the first 3 lasers dynamics are plotted (dashed and dotted lines). The mean dynamics is plotted in continuous black. The parameters $\frac{\delta_i-\left\langle \delta\right\rangle }{\left\langle \delta\right\rangle}$ are independent Gaussian variables of zero mean and standard deviation $1e-3$ with $ \left\langle \delta\right\rangle = 1.2045$, $k=0.7$, $A=1$, $\alpha=2$, $\gamma=4*10^{-3}$, $\epsilon=10^{-4}$. Black dot: intersection between the two slow manifold branches. The parabola $\Sigma_y$ and the straight line $\Sigma_x$ constitute the critical manifold calculated for a single laser with parameter $\Delta$.}
	
	\label{numerics}
\end{figure}

Beyond stationary states, much insight can be gained by studying the stability of the critical manifold. Defining the mean carrier density $Y = \frac{1}{N}\sum\limits_{i=1}^N y_i$ we derive the following rate equations:
\vspace{-0.3cm}
\begin{eqnarray} 
\dot{X} &=& - X + \frac{1}{N}\sum\limits_{i=1}^N x_i y_i \;  \label{eqa2} \\
\dot{Y} &=& \gamma (\Delta - Y + k (w + f(X)) - \frac{1}{N}\sum\limits_{i=1}^N x_i y_i) \;
\label{eqb2}\\
\dot{w} &=& -\epsilon (w + f(X) ).\;  \label{eqc2}
\end{eqnarray}
From Eq. \ref{eqa2}, we have that $\dot{X} = 0 \Leftrightarrow X = \frac{1}{N}\sum\limits_{i=1}^N x_i y_i$. The critical manifold is solution of $\Delta - Y + k w + k f(X) - X=0$. It is clear that $\dot{X} = 0$ is satisfied either if all lasers are off: $x_i=0$ $\forall i$, which gives 
$Y_w = \Delta + k w$ , or if all lasers are on: $y_i=1$ $\forall i$, so that $Y=1$. This provides two of the 1D branches of the critical manifold of the full network. These curves are defined by exactly the same equations as for $\Sigma$, but where all the variables and parameters are replaced by their corresponding mean values. 
To analyze the critical manifold in the general case, we parameterize it by the
set $I$ of switched ON lasers and we introduce the new variable $X_I =\frac1N\sum\limits_{i\in I}x_i$ and parameter $\Delta_I = \frac1N\sum\limits_{i\in I}\delta_i$. We find that:
\[S_I = \{ (x^I_i(w),y^I_i(w),w),\ i=1\cdots N,\ w\in\mathbb R\}\]
with
\[
(x_i^I(w),y^I_i(w)) = \left\lbrace  
\begin{aligned}
(0,\delta_i+k(w+f(X_I(w)))),&\quad \forall i\notin I,\\
(\delta_i-1+k(w+f(X_I(w))),1)&\quad \forall i\in I,
\end{aligned}\right.
\]
where $X_I(w)$ is implicitly defined by
\begin{equation}
X_I(w) = \frac{N_+}{N}(k(w+f(X_I(w)))-1)+\Delta_I. 
\end{equation}
The critical manifold of the mean-field coupled network thus consists of $2^N$ components: $S=\cup_{I\subset[1,N]}S_I$. 
Apart from the scaling factor $\frac{N_+}{N}$, the structure of the critical manifold is a bundle of 1D branches $S_I$ which, at zero order in $z$, closely resembles that of the $N=1$ case except for the OFF part. As for equilibria, the stability of $S_I$ can be determined analytically assuming that all lasers are similar enough $\delta_i = \Delta+z\eta_i, z\ll1$ \cite{si}. It turns out that the stability of $S_I$, at zero order $z$, differs from that of a single "mean" laser (with control parameter $\Delta$ and which incorporates the proportion $\frac{N_+}{N}$) by the destabilizing effect of the off part due to global coupling.

\begin{figure}[t]
	\includegraphics[width=0.45\textwidth]{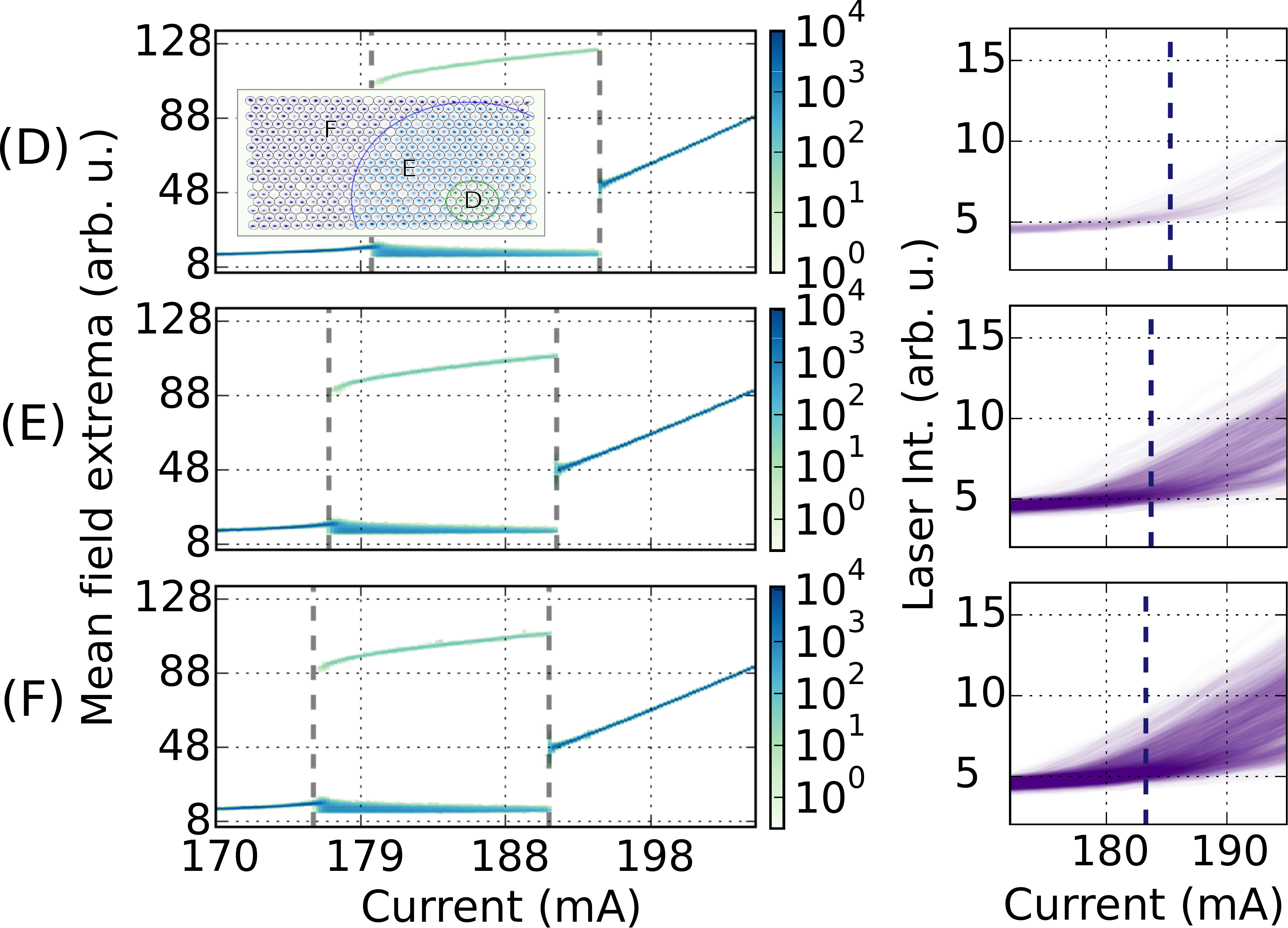}
	\caption{Bifurcation diagrams for the slow manifold of three networks of different sizes (D:19, E:251, F:451). Left: The dashed lines indicate the bifurcations from and to the only stable fixed point. Right: transcritical laser bifurcation curves for the uncoupled elements, the dashed line indicates the ensemble average of the threshold current values $\langle I_{th}\rangle$.\label{fig:growing}}
\end{figure}

Thus, as the quenched disorder $\delta_i$ is not averaged in the limit $N\to\infty$, a truly mean field limit cannot be established as an ODE. However, due to the splitting of the timescales, most of the motion takes place along the critical manifold leading to an effective low dimensional dynamics very similar to that of a single element. In \fig{numerics}, we plot the numerical mean field trajectory together with the critical manifold of an average laser. The slow evolution of different nodes is perfectly synchronized, even if some elements may be on different branches of the slow manifold (as in the experimental observation of \fig{fig:setup} d)). However, the individual trajectories differ in the fast part of the dynamics, which is transversal to the slow manifold. This is clear on the bottom of \fig{numerics} which shows a time trace of the mean field together with the variance of the $x_i$. In absence of noise, the distribution of the $x_i$ tendis to a Dirac function whenever the system is close to the critical manifold with a much broader distribution when the system switches branch.

Finally, the dynamics for $z\ll1, N<\infty$ stays in a tube around the dynamics for $z=0$ in which case there is no disorder and thus the dynamics is exactly that of the isolated laser. This implies that the MMO and the chaotic behaviors (for $z=0$) are robust on finite time intervals when $N$ grows to infinity. This is exemplified in \fig{numerics} which shows a chaotic trajectory in the case $z\ll1, N=10^4$.

We assess the impact of $\Delta$ experimentally by measuring the total intensity for different population sizes (\fig{fig:growing}). All parameters are  constant and the iris is opened to include a larger and larger population. For each network size, the total amount of light sent to the reinjection detector is scaled to keep the coupling parameter constant. We show the bifurcation diagrams of networks of 19 (D), 251 (E) and 451 (F) nodes on \fig{fig:growing}. Similar sequences are observed, although for different values of the control parameter. The distributions of the uncoupled laser emission thresholds are shown on the right column. The 451 and 251 elements networks are very similar but the 19 element one differs markedly. As expected from theory, this hints at  $\Delta$ as control parameter for the network.

We demonstrate this convergence by measuring the current value at which some prescribed dynamics takes place for different populations. On \fig{fig:convergence}, we plot the current value \(\mathcal{I}_s\) at which the network returns to a stable fixed point after undergoing the sequence of bifurcations described earlier, as a function of the average threshold current of the sub-population. The size and color of each marker indicate the size of the network. The error bars are estimates of the measurement error. Smaller networks are disperse but larger networks converge towards the same point in this $(<I_{th}>,\mathcal{I}_s)$ space. The dispersion of the measurements around a straight line results from the scaling of the bifurcation parameter $\Delta=\frac{I_0-I_t}{<I_{th}>-I_t}$ where $I_t$ is the transparency current (assumed to be equal for all devices).

\begin{figure}[t]
	\setlength{\unitlength}{1cm}
	\begin{picture}(8,4)
	\put(0,0){\includegraphics[width=0.45\textwidth]{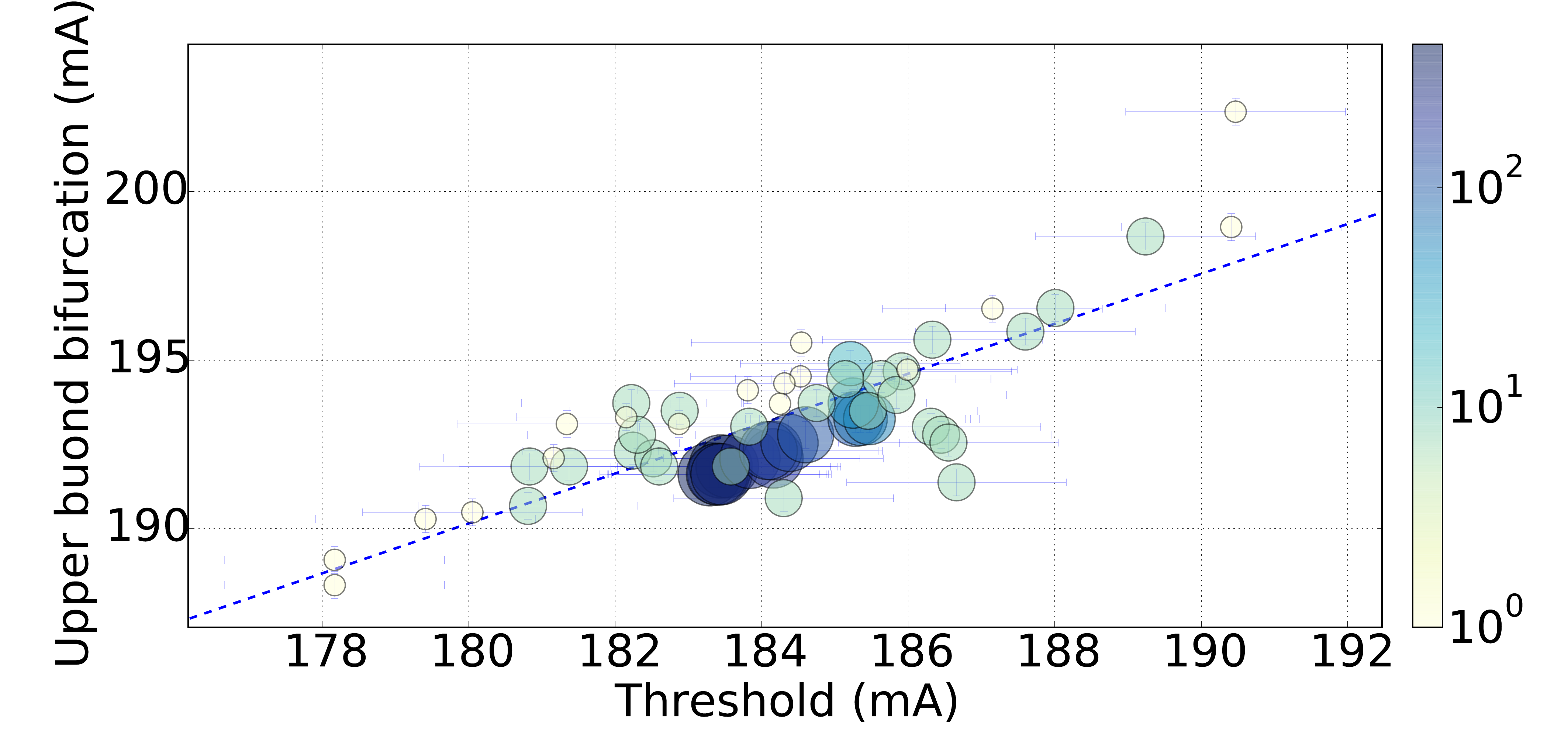}}
	\put(1.1,2.){\includegraphics[width=0.18\textwidth]{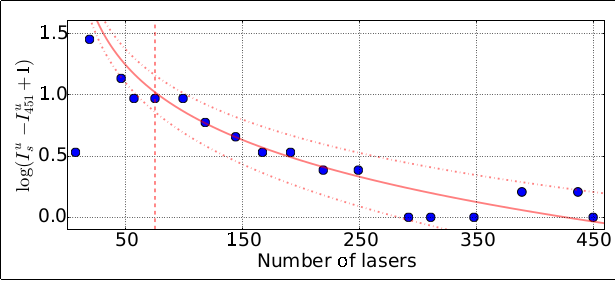}}
	\end{picture}
	\caption{\label{fig3}Upper bifurcation parameter value depending on average laser threshold value for increasing sample size. Smaller samples (lighter blue) are distributed along a straight line. When the sample grows (darker blue) the bifurcation parameter converges to a well defined value. Inset: convergence of the empirical mean towards its final value for the largest population and fit in $1/\sqrt{N}$.
		\label{fig:convergence}}
	\label{fig:convergence}
\end{figure}

Summarizing, we have reported experimental observations of mixed mode oscillations and spiking in a mean field coupled network of hundreds of semiconductor lasers chaotically spiking and coupled through nonlinear optoelectronic feedback. A transport equation for the probability density $p(t,x,y,w,\delta)$ of the limit laser for $N \longrightarrow \infty$ involves the full distribution of the $\delta_i$, which shows that the mean field cannot be described with an ODE. These phenomena result from the slow-fast nature of the system. Through the stability analysis of the critical manifold, we demonstrate that the network experiences an effectively low-dimensional dynamics even when the fast dynamics of the nodes is not synchronized. These experimental observations show that the results are robust with respect to some amount of disorder in the couplings. Thanks to the relative simplicity of the experimental platform, we expect that the present results open several research avenues, specifically on the role of noise in coupled slow-fast systems and on networks of networks.

\begin{acknowledgments}
	The authors acknowledge support of Région Provence Alpes Côte d'Azur through project SYNCOP (DEB 15-1383 and DEB 15-1376). FM thanks CNRS for funding his stay at Institut de Physique de Nice. This work was conducted within the framework of the project OPTIMAL granted by the European Union by means of the Fond Européen de développement regional, FEDER. We thank Dr. Otti d'Huys for many insightful discussions.

\end{acknowledgments}

\bibliography{meanfield}

\begin{thebibliography}{34}
\expandafter\ifx\csname natexlab\endcsname\relax\def\natexlab#1{#1}\fi
\expandafter\ifx\csname bibnamefont\endcsname\relax
  \def\bibnamefont#1{#1}\fi
\expandafter\ifx\csname bibfnamefont\endcsname\relax
  \def\bibfnamefont#1{#1}\fi
\expandafter\ifx\csname citenamefont\endcsname\relax
  \def\citenamefont#1{#1}\fi
\expandafter\ifx\csname url\endcsname\relax
  \def\url#1{\texttt{#1}}\fi
\expandafter\ifx\csname urlprefix\endcsname\relax\def\urlprefix{URL }\fi
\providecommand{\bibinfo}[2]{#2}
\providecommand{\eprint}[2][]{\url{#2}}

\bibitem[{\citenamefont{N{\'e}da et~al.}(2000)\citenamefont{N{\'e}da, Ravasz,
  Vicsek, Brechet, and Barab{\'a}si}}]{neda2000physics}
\bibinfo{author}{\bibfnamefont{Z.}~\bibnamefont{N{\'e}da}},
  \bibinfo{author}{\bibfnamefont{E.}~\bibnamefont{Ravasz}},
  \bibinfo{author}{\bibfnamefont{T.}~\bibnamefont{Vicsek}},
  \bibinfo{author}{\bibfnamefont{Y.}~\bibnamefont{Brechet}}, \bibnamefont{and}
  \bibinfo{author}{\bibfnamefont{A.-L.} \bibnamefont{Barab{\'a}si}},
  \bibinfo{journal}{Physical Review E} \textbf{\bibinfo{volume}{61}},
  \bibinfo{pages}{6987} (\bibinfo{year}{2000}).

\bibitem[{\citenamefont{Kuramoto}(2012)}]{kuramoto2012chemical}
\bibinfo{author}{\bibfnamefont{Y.}~\bibnamefont{Kuramoto}},
  \emph{\bibinfo{title}{Chemical oscillations, waves, and turbulence}},
  vol.~\bibinfo{volume}{19} (\bibinfo{publisher}{Springer Science \& Business
  Media}, \bibinfo{year}{2012}).

\bibitem[{\citenamefont{Strogatz}(2000)}]{strogatz2000kuramoto}
\bibinfo{author}{\bibfnamefont{S.~H.} \bibnamefont{Strogatz}},
  \bibinfo{journal}{Physica D: Nonlinear Phenomena}
  \textbf{\bibinfo{volume}{143}}, \bibinfo{pages}{1} (\bibinfo{year}{2000}).

\bibitem[{\citenamefont{Jiruska et~al.}(2013)\citenamefont{Jiruska, De~Curtis,
  Jefferys, Schevon, Schiff, and Schindler}}]{jiruska2013synchronization}
\bibinfo{author}{\bibfnamefont{P.}~\bibnamefont{Jiruska}},
  \bibinfo{author}{\bibfnamefont{M.}~\bibnamefont{De~Curtis}},
  \bibinfo{author}{\bibfnamefont{J.~G.} \bibnamefont{Jefferys}},
  \bibinfo{author}{\bibfnamefont{C.~A.} \bibnamefont{Schevon}},
  \bibinfo{author}{\bibfnamefont{S.~J.} \bibnamefont{Schiff}},
  \bibnamefont{and}
  \bibinfo{author}{\bibfnamefont{K.}~\bibnamefont{Schindler}},
  \bibinfo{journal}{The Journal of physiology} \textbf{\bibinfo{volume}{591}},
  \bibinfo{pages}{787} (\bibinfo{year}{2013}).

\bibitem[{\citenamefont{Mirollo and
  Strogatz}(1990)}]{mirollo1990synchronization}
\bibinfo{author}{\bibfnamefont{R.~E.} \bibnamefont{Mirollo}} \bibnamefont{and}
  \bibinfo{author}{\bibfnamefont{S.~H.} \bibnamefont{Strogatz}},
  \bibinfo{journal}{SIAM Journal on Applied Mathematics}
  \textbf{\bibinfo{volume}{50}}, \bibinfo{pages}{1645} (\bibinfo{year}{1990}).

\bibitem[{\citenamefont{Watanabe and Strogatz}(1994)}]{watanabe1994constants}
\bibinfo{author}{\bibfnamefont{S.}~\bibnamefont{Watanabe}} \bibnamefont{and}
  \bibinfo{author}{\bibfnamefont{S.~H.} \bibnamefont{Strogatz}},
  \bibinfo{journal}{Physica D: Nonlinear Phenomena}
  \textbf{\bibinfo{volume}{74}}, \bibinfo{pages}{197} (\bibinfo{year}{1994}).

\bibitem[{\citenamefont{Zillmer et~al.}(2006)\citenamefont{Zillmer, Livi,
  Politi, and Torcini}}]{zillmer2006desynchronization}
\bibinfo{author}{\bibfnamefont{R.}~\bibnamefont{Zillmer}},
  \bibinfo{author}{\bibfnamefont{R.}~\bibnamefont{Livi}},
  \bibinfo{author}{\bibfnamefont{A.}~\bibnamefont{Politi}}, \bibnamefont{and}
  \bibinfo{author}{\bibfnamefont{A.}~\bibnamefont{Torcini}},
  \bibinfo{journal}{Physical Review E} \textbf{\bibinfo{volume}{74}},
  \bibinfo{pages}{036203} (\bibinfo{year}{2006}).

\bibitem[{\citenamefont{Olmi et~al.}(2014)\citenamefont{Olmi, Navas,
  Boccaletti, and Torcini}}]{olmi2014hysteretic}
\bibinfo{author}{\bibfnamefont{S.}~\bibnamefont{Olmi}},
  \bibinfo{author}{\bibfnamefont{A.}~\bibnamefont{Navas}},
  \bibinfo{author}{\bibfnamefont{S.}~\bibnamefont{Boccaletti}},
  \bibnamefont{and} \bibinfo{author}{\bibfnamefont{A.}~\bibnamefont{Torcini}},
  \bibinfo{journal}{Physical Review E} \textbf{\bibinfo{volume}{90}},
  \bibinfo{pages}{042905} (\bibinfo{year}{2014}).

\bibitem[{\citenamefont{Kotani et~al.}(2014)\citenamefont{Kotani, Yamaguchi,
  Yoshida, Jimbo, and Ermentrout}}]{kotani2014population}
\bibinfo{author}{\bibfnamefont{K.}~\bibnamefont{Kotani}},
  \bibinfo{author}{\bibfnamefont{I.}~\bibnamefont{Yamaguchi}},
  \bibinfo{author}{\bibfnamefont{L.}~\bibnamefont{Yoshida}},
  \bibinfo{author}{\bibfnamefont{Y.}~\bibnamefont{Jimbo}}, \bibnamefont{and}
  \bibinfo{author}{\bibfnamefont{G.~B.} \bibnamefont{Ermentrout}},
  \bibinfo{journal}{Journal of The Royal Society Interface}
  \textbf{\bibinfo{volume}{11}}, \bibinfo{pages}{20140058}
  (\bibinfo{year}{2014}).

\bibitem[{\citenamefont{Montbri{\'o} et~al.}(2015)\citenamefont{Montbri{\'o},
  Paz{\'o}, and Roxin}}]{montbrio2015macroscopic}
\bibinfo{author}{\bibfnamefont{E.}~\bibnamefont{Montbri{\'o}}},
  \bibinfo{author}{\bibfnamefont{D.}~\bibnamefont{Paz{\'o}}}, \bibnamefont{and}
  \bibinfo{author}{\bibfnamefont{A.}~\bibnamefont{Roxin}},
  \bibinfo{journal}{Physical Review X} \textbf{\bibinfo{volume}{5}},
  \bibinfo{pages}{021028} (\bibinfo{year}{2015}).

\bibitem[{\citenamefont{Paz{\'o} and
  Montbri{\'o}}(2016)}]{pazo2016quasiperiodic}
\bibinfo{author}{\bibfnamefont{D.}~\bibnamefont{Paz{\'o}}} \bibnamefont{and}
  \bibinfo{author}{\bibfnamefont{E.}~\bibnamefont{Montbri{\'o}}},
  \bibinfo{journal}{Physical review letters} \textbf{\bibinfo{volume}{116}},
  \bibinfo{pages}{238101} (\bibinfo{year}{2016}).

\bibitem[{\citenamefont{Ott and Antonsen}(2008)}]{ott2008low}
\bibinfo{author}{\bibfnamefont{E.}~\bibnamefont{Ott}} \bibnamefont{and}
  \bibinfo{author}{\bibfnamefont{T.~M.} \bibnamefont{Antonsen}},
  \bibinfo{journal}{Chaos: An Interdisciplinary Journal of Nonlinear Science}
  \textbf{\bibinfo{volume}{18}}, \bibinfo{pages}{037113}
  (\bibinfo{year}{2008}).

\bibitem[{\citenamefont{Ott and Antonsen}(2009)}]{ott2009long}
\bibinfo{author}{\bibfnamefont{E.}~\bibnamefont{Ott}} \bibnamefont{and}
  \bibinfo{author}{\bibfnamefont{T.~M.} \bibnamefont{Antonsen}},
  \bibinfo{journal}{Chaos: An interdisciplinary journal of nonlinear science}
  \textbf{\bibinfo{volume}{19}}, \bibinfo{pages}{023117}
  (\bibinfo{year}{2009}).

\bibitem[{\citenamefont{Nixon et~al.}(2011)\citenamefont{Nixon, Friedman,
  Ronen, Friesem, Davidson, and Kanter}}]{nixon2011synchronized}
\bibinfo{author}{\bibfnamefont{M.}~\bibnamefont{Nixon}},
  \bibinfo{author}{\bibfnamefont{M.}~\bibnamefont{Friedman}},
  \bibinfo{author}{\bibfnamefont{E.}~\bibnamefont{Ronen}},
  \bibinfo{author}{\bibfnamefont{A.~A.} \bibnamefont{Friesem}},
  \bibinfo{author}{\bibfnamefont{N.}~\bibnamefont{Davidson}}, \bibnamefont{and}
  \bibinfo{author}{\bibfnamefont{I.}~\bibnamefont{Kanter}},
  \bibinfo{journal}{Physical review letters} \textbf{\bibinfo{volume}{106}},
  \bibinfo{pages}{223901} (\bibinfo{year}{2011}).

\bibitem[{\citenamefont{Nixon et~al.}(2012)\citenamefont{Nixon, Fridman, Ronen,
  Friesem, Davidson, and Kanter}}]{nixon2012controlling}
\bibinfo{author}{\bibfnamefont{M.}~\bibnamefont{Nixon}},
  \bibinfo{author}{\bibfnamefont{M.}~\bibnamefont{Fridman}},
  \bibinfo{author}{\bibfnamefont{E.}~\bibnamefont{Ronen}},
  \bibinfo{author}{\bibfnamefont{A.~A.} \bibnamefont{Friesem}},
  \bibinfo{author}{\bibfnamefont{N.}~\bibnamefont{Davidson}}, \bibnamefont{and}
  \bibinfo{author}{\bibfnamefont{I.}~\bibnamefont{Kanter}},
  \bibinfo{journal}{Physical review letters} \textbf{\bibinfo{volume}{108}},
  \bibinfo{pages}{214101} (\bibinfo{year}{2012}).

\bibitem[{\citenamefont{Selmi et~al.}(2014)\citenamefont{Selmi, Braive,
  Beaudoin, Sagnes, Kuszelewicz, and Barbay}}]{selmi2014relative}
\bibinfo{author}{\bibfnamefont{F.}~\bibnamefont{Selmi}},
  \bibinfo{author}{\bibfnamefont{R.}~\bibnamefont{Braive}},
  \bibinfo{author}{\bibfnamefont{G.}~\bibnamefont{Beaudoin}},
  \bibinfo{author}{\bibfnamefont{I.}~\bibnamefont{Sagnes}},
  \bibinfo{author}{\bibfnamefont{R.}~\bibnamefont{Kuszelewicz}},
  \bibnamefont{and} \bibinfo{author}{\bibfnamefont{S.}~\bibnamefont{Barbay}},
  \bibinfo{journal}{Physical review letters} \textbf{\bibinfo{volume}{112}},
  \bibinfo{pages}{183902} (\bibinfo{year}{2014}).

\bibitem[{\citenamefont{Hurtado and Javaloyes}(2015)}]{hurtado2015controllable}
\bibinfo{author}{\bibfnamefont{A.}~\bibnamefont{Hurtado}} \bibnamefont{and}
  \bibinfo{author}{\bibfnamefont{J.}~\bibnamefont{Javaloyes}},
  \bibinfo{journal}{Applied Physics Letters} \textbf{\bibinfo{volume}{107}},
  \bibinfo{pages}{241103} (\bibinfo{year}{2015}).

\bibitem[{\citenamefont{Sorrentino et~al.}(2015)\citenamefont{Sorrentino,
  Quintero-Quiroz, Aragoneses, Torrent, and Masoller}}]{sorrentino2015effects}
\bibinfo{author}{\bibfnamefont{T.}~\bibnamefont{Sorrentino}},
  \bibinfo{author}{\bibfnamefont{C.}~\bibnamefont{Quintero-Quiroz}},
  \bibinfo{author}{\bibfnamefont{A.}~\bibnamefont{Aragoneses}},
  \bibinfo{author}{\bibfnamefont{M.}~\bibnamefont{Torrent}}, \bibnamefont{and}
  \bibinfo{author}{\bibfnamefont{C.}~\bibnamefont{Masoller}},
  \bibinfo{journal}{Optics express} \textbf{\bibinfo{volume}{23}},
  \bibinfo{pages}{5571} (\bibinfo{year}{2015}).

\bibitem[{\citenamefont{Mesaritakis et~al.}(2016)\citenamefont{Mesaritakis,
  Kapsalis, Bogris, and Syvridis}}]{mesaritakis2016artificial}
\bibinfo{author}{\bibfnamefont{C.}~\bibnamefont{Mesaritakis}},
  \bibinfo{author}{\bibfnamefont{A.}~\bibnamefont{Kapsalis}},
  \bibinfo{author}{\bibfnamefont{A.}~\bibnamefont{Bogris}}, \bibnamefont{and}
  \bibinfo{author}{\bibfnamefont{D.}~\bibnamefont{Syvridis}},
  \bibinfo{journal}{Scientific reports} \textbf{\bibinfo{volume}{6}},
  \bibinfo{pages}{39317} (\bibinfo{year}{2016}).

\bibitem[{\citenamefont{Prucnal et~al.}(2016)\citenamefont{Prucnal, Shastri,
  de~Lima, Nahmias, and Tait}}]{prucnal2016recent}
\bibinfo{author}{\bibfnamefont{P.~R.} \bibnamefont{Prucnal}},
  \bibinfo{author}{\bibfnamefont{B.~J.} \bibnamefont{Shastri}},
  \bibinfo{author}{\bibfnamefont{T.~F.} \bibnamefont{de~Lima}},
  \bibinfo{author}{\bibfnamefont{M.~A.} \bibnamefont{Nahmias}},
  \bibnamefont{and} \bibinfo{author}{\bibfnamefont{A.~N.} \bibnamefont{Tait}},
  \bibinfo{journal}{Advances in Optics and Photonics}
  \textbf{\bibinfo{volume}{8}}, \bibinfo{pages}{228} (\bibinfo{year}{2016}).

\bibitem[{\citenamefont{Dolcemascolo et~al.}(2018)\citenamefont{Dolcemascolo,
  Garbin, Peyce, Veltz, and Barland}}]{dolcemascolo2018resonator}
\bibinfo{author}{\bibfnamefont{A.}~\bibnamefont{Dolcemascolo}},
  \bibinfo{author}{\bibfnamefont{B.}~\bibnamefont{Garbin}},
  \bibinfo{author}{\bibfnamefont{B.}~\bibnamefont{Peyce}},
  \bibinfo{author}{\bibfnamefont{R.}~\bibnamefont{Veltz}}, \bibnamefont{and}
  \bibinfo{author}{\bibfnamefont{S.}~\bibnamefont{Barland}},
  \bibinfo{journal}{Physical Review E} \textbf{\bibinfo{volume}{98}},
  \bibinfo{pages}{062211} (\bibinfo{year}{2018}).

\bibitem[{\citenamefont{Garbin et~al.}(2015)\citenamefont{Garbin, Javaloyes,
  Tissoni, and Barland}}]{garbin2015topological}
\bibinfo{author}{\bibfnamefont{B.}~\bibnamefont{Garbin}},
  \bibinfo{author}{\bibfnamefont{J.}~\bibnamefont{Javaloyes}},
  \bibinfo{author}{\bibfnamefont{G.}~\bibnamefont{Tissoni}}, \bibnamefont{and}
  \bibinfo{author}{\bibfnamefont{S.}~\bibnamefont{Barland}},
  \bibinfo{journal}{Nature communications} \textbf{\bibinfo{volume}{6}},
  \bibinfo{pages}{5915} (\bibinfo{year}{2015}).

\bibitem[{\citenamefont{Romeira et~al.}(2016)\citenamefont{Romeira, Av{\'{o}},
  Figueiredo, Barland, and Javaloyes}}]{romeira2016regenerative}
\bibinfo{author}{\bibfnamefont{B.}~\bibnamefont{Romeira}},
  \bibinfo{author}{\bibfnamefont{R.}~\bibnamefont{Av{\'{o}}}},
  \bibinfo{author}{\bibfnamefont{J.~M.} \bibnamefont{Figueiredo}},
  \bibinfo{author}{\bibfnamefont{S.}~\bibnamefont{Barland}}, \bibnamefont{and}
  \bibinfo{author}{\bibfnamefont{J.}~\bibnamefont{Javaloyes}},
  \bibinfo{journal}{Scientific reports} \textbf{\bibinfo{volume}{6}}
  (\bibinfo{year}{2016}).

\bibitem[{\citenamefont{Terrien et~al.}(2018)\citenamefont{Terrien, Krauskopf,
  Broderick, Braive, Beaudoin, Sagnes, and Barbay}}]{terrien2018pulse}
\bibinfo{author}{\bibfnamefont{S.}~\bibnamefont{Terrien}},
  \bibinfo{author}{\bibfnamefont{B.}~\bibnamefont{Krauskopf}},
  \bibinfo{author}{\bibfnamefont{N.~G.} \bibnamefont{Broderick}},
  \bibinfo{author}{\bibfnamefont{R.}~\bibnamefont{Braive}},
  \bibinfo{author}{\bibfnamefont{G.}~\bibnamefont{Beaudoin}},
  \bibinfo{author}{\bibfnamefont{I.}~\bibnamefont{Sagnes}}, \bibnamefont{and}
  \bibinfo{author}{\bibfnamefont{S.}~\bibnamefont{Barbay}},
  \bibinfo{journal}{Optics Letters} \textbf{\bibinfo{volume}{43}},
  \bibinfo{pages}{3013} (\bibinfo{year}{2018}).

\bibitem[{\citenamefont{Yacomotti et~al.}(2002)\citenamefont{Yacomotti,
  Mindlin, Giudici, Balle, Barland, and Tredicce}}]{yacomotti2002coupled}
\bibinfo{author}{\bibfnamefont{A.~M.} \bibnamefont{Yacomotti}},
  \bibinfo{author}{\bibfnamefont{G.~B.} \bibnamefont{Mindlin}},
  \bibinfo{author}{\bibfnamefont{M.}~\bibnamefont{Giudici}},
  \bibinfo{author}{\bibfnamefont{S.}~\bibnamefont{Balle}},
  \bibinfo{author}{\bibfnamefont{S.}~\bibnamefont{Barland}}, \bibnamefont{and}
  \bibinfo{author}{\bibfnamefont{J.}~\bibnamefont{Tredicce}},
  \bibinfo{journal}{Physical Review E} \textbf{\bibinfo{volume}{66}},
  \bibinfo{pages}{036227} (\bibinfo{year}{2002}).

\bibitem[{\citenamefont{Kelleher et~al.}(2010)\citenamefont{Kelleher, Bonatto,
  Skoda, Hegarty, and Huyet}}]{kelleher2010excitation}
\bibinfo{author}{\bibfnamefont{B.}~\bibnamefont{Kelleher}},
  \bibinfo{author}{\bibfnamefont{C.}~\bibnamefont{Bonatto}},
  \bibinfo{author}{\bibfnamefont{P.}~\bibnamefont{Skoda}},
  \bibinfo{author}{\bibfnamefont{S.}~\bibnamefont{Hegarty}}, \bibnamefont{and}
  \bibinfo{author}{\bibfnamefont{G.}~\bibnamefont{Huyet}},
  \bibinfo{journal}{Physical Review E} \textbf{\bibinfo{volume}{81}},
  \bibinfo{pages}{036204} (\bibinfo{year}{2010}).

\bibitem[{\citenamefont{Van~Vaerenbergh
  et~al.}(2012)\citenamefont{Van~Vaerenbergh, Fiers, Mechet, Spuesens, Kumar,
  Morthier, Schrauwen, Dambre, and Bienstman}}]{van2012cascadable}
\bibinfo{author}{\bibfnamefont{T.}~\bibnamefont{Van~Vaerenbergh}},
  \bibinfo{author}{\bibfnamefont{M.}~\bibnamefont{Fiers}},
  \bibinfo{author}{\bibfnamefont{P.}~\bibnamefont{Mechet}},
  \bibinfo{author}{\bibfnamefont{T.}~\bibnamefont{Spuesens}},
  \bibinfo{author}{\bibfnamefont{R.}~\bibnamefont{Kumar}},
  \bibinfo{author}{\bibfnamefont{G.}~\bibnamefont{Morthier}},
  \bibinfo{author}{\bibfnamefont{B.}~\bibnamefont{Schrauwen}},
  \bibinfo{author}{\bibfnamefont{J.}~\bibnamefont{Dambre}}, \bibnamefont{and}
  \bibinfo{author}{\bibfnamefont{P.}~\bibnamefont{Bienstman}},
  \bibinfo{journal}{Optics express} \textbf{\bibinfo{volume}{20}},
  \bibinfo{pages}{20292} (\bibinfo{year}{2012}).

\bibitem[{\citenamefont{Deng et~al.}(2017)\citenamefont{Deng, Robertson, and
  Hurtado}}]{deng2017controlled}
\bibinfo{author}{\bibfnamefont{T.}~\bibnamefont{Deng}},
  \bibinfo{author}{\bibfnamefont{J.}~\bibnamefont{Robertson}},
  \bibnamefont{and} \bibinfo{author}{\bibfnamefont{A.}~\bibnamefont{Hurtado}},
  \bibinfo{journal}{IEEE Journal of Selected Topics in Quantum Electronics}
  \textbf{\bibinfo{volume}{23}}, \bibinfo{pages}{1} (\bibinfo{year}{2017}).

\bibitem[{\citenamefont{Belykh et~al.}(2005)\citenamefont{Belykh, de~Lange, and
  Hasler}}]{belykh2005synchronization}
\bibinfo{author}{\bibfnamefont{I.}~\bibnamefont{Belykh}},
  \bibinfo{author}{\bibfnamefont{E.}~\bibnamefont{de~Lange}}, \bibnamefont{and}
  \bibinfo{author}{\bibfnamefont{M.}~\bibnamefont{Hasler}},
  \bibinfo{journal}{Physical review letters} \textbf{\bibinfo{volume}{94}},
  \bibinfo{pages}{188101} (\bibinfo{year}{2005}).

\bibitem[{\citenamefont{Al-Naimee et~al.}(2009)\citenamefont{Al-Naimee, Marino,
  Ciszak, Meucci, and Arecchi}}]{al2009chaotic}
\bibinfo{author}{\bibfnamefont{K.}~\bibnamefont{Al-Naimee}},
  \bibinfo{author}{\bibfnamefont{F.}~\bibnamefont{Marino}},
  \bibinfo{author}{\bibfnamefont{M.}~\bibnamefont{Ciszak}},
  \bibinfo{author}{\bibfnamefont{R.}~\bibnamefont{Meucci}}, \bibnamefont{and}
  \bibinfo{author}{\bibfnamefont{F.~T.} \bibnamefont{Arecchi}},
  \bibinfo{journal}{New Journal of Physics} \textbf{\bibinfo{volume}{11}},
  \bibinfo{pages}{073022} (\bibinfo{year}{2009}).

\bibitem[{\citenamefont{Marino et~al.}(2011)\citenamefont{Marino, Ciszak,
  Abdalah, Al-Naimee, Meucci, and Arecchi}}]{marino2011mixed}
\bibinfo{author}{\bibfnamefont{F.}~\bibnamefont{Marino}},
  \bibinfo{author}{\bibfnamefont{M.}~\bibnamefont{Ciszak}},
  \bibinfo{author}{\bibfnamefont{S.}~\bibnamefont{Abdalah}},
  \bibinfo{author}{\bibfnamefont{K.}~\bibnamefont{Al-Naimee}},
  \bibinfo{author}{\bibfnamefont{R.}~\bibnamefont{Meucci}}, \bibnamefont{and}
  \bibinfo{author}{\bibfnamefont{F.}~\bibnamefont{Arecchi}},
  \bibinfo{journal}{Physical Review E} \textbf{\bibinfo{volume}{84}},
  \bibinfo{pages}{047201} (\bibinfo{year}{2011}).

\bibitem[{si()}]{si}
\bibinfo{note}{See Supplemental Material}.

\bibitem[{\citenamefont{Al-Naimee et~al.}(2010)\citenamefont{Al-Naimee, Marino,
  Ciszak, Abdalah, Meucci, and Arecchi}}]{al2010excitability}
\bibinfo{author}{\bibfnamefont{K.}~\bibnamefont{Al-Naimee}},
  \bibinfo{author}{\bibfnamefont{F.}~\bibnamefont{Marino}},
  \bibinfo{author}{\bibfnamefont{M.}~\bibnamefont{Ciszak}},
  \bibinfo{author}{\bibfnamefont{S.}~\bibnamefont{Abdalah}},
  \bibinfo{author}{\bibfnamefont{R.}~\bibnamefont{Meucci}}, \bibnamefont{and}
  \bibinfo{author}{\bibfnamefont{F.}~\bibnamefont{Arecchi}},
  \bibinfo{journal}{The European Physical Journal D}
  \textbf{\bibinfo{volume}{58}}, \bibinfo{pages}{187} (\bibinfo{year}{2010}).

\bibitem[{\citenamefont{Desroches et~al.}(2012)\citenamefont{Desroches,
  Guckenheimer, Krauskopf, Kuehn, Osinga, and
  Wechselberger}}]{desroches2012mixed}
\bibinfo{author}{\bibfnamefont{M.}~\bibnamefont{Desroches}},
  \bibinfo{author}{\bibfnamefont{J.}~\bibnamefont{Guckenheimer}},
  \bibinfo{author}{\bibfnamefont{B.}~\bibnamefont{Krauskopf}},
  \bibinfo{author}{\bibfnamefont{C.}~\bibnamefont{Kuehn}},
  \bibinfo{author}{\bibfnamefont{H.~M.} \bibnamefont{Osinga}},
  \bibnamefont{and}
  \bibinfo{author}{\bibfnamefont{M.}~\bibnamefont{Wechselberger}},
  \bibinfo{journal}{Siam Review} \textbf{\bibinfo{volume}{54}},
  \bibinfo{pages}{211} (\bibinfo{year}{2012}).

\end{thebibliography}

\end{document}

\title{Effective low-dimensional dynamics of a mean-field coupled network of slow-fast spiking lasers - Supplemental information}

    \author{A.~Dolcemascolo}
      \author{A.~Miazek}
        \affiliation{Universit\'e C\^ote d'Azur, CNRS, INPHYNI, 1361 route des lucioles 06560 Valbonne, France}
	  \author{R.~Veltz}
	    \affiliation{Inria Sophia Antipolis, MathNeuro Team, 2004 route des Lucioles - BP93, 06902 Sophia         Antipolis, France}
	      \author{F.~Marino}
	      >>\affiliation{CNR-Istituto Nazionale di Ottica and INFN, Sez. di Firenze, Via Sansone 1, I-50019 Sesto     Fiorentino (FI), Italy}
	        \author{S. Barland}
		  \affiliation{Universit\'e C\^ote d'Azur, CNRS, INPHYNI, 1361 route des lucioles 06560 Valbonne, France}

\maketitle

\section{Model scaling}

The single laser with optoelectronic feedback can be modelled with the following system:

\begin{eqnarray}
\dot{s} & = & \big[g(n-n_t)-\gamma_0\big]s \\\
\dot{n} & = & \frac{I_0+kI}{ev} - \gamma_cn - g(n-n_t)s \\\
\dot{I} & = & -\gamma_fI+\dot{f_f(s)} 
\end{eqnarray}

After proper scaling $x=\frac{g}{\gamma_c s}$, $y=\frac {g}{\gamma_0}(n-n_t)$ and $w=I-f_f(\frac{\gamma_c}{g}x)$, we end up with the following system for a fully connected network for $i=1,\cdots ,N$:

\begin{eqnarray}\label{eq:model}
\dot{x_i} & = & x_i(y_i-1) \;  \label{eq:model1} \\
\dot{y_i} & = & \gamma(\delta_i-y_i+k(w+f(X)) - x_iy_i) \;  \label{eq:model2} \\
\dot{w} & = & -\epsilon(w + f(X)) \;  \label{eq:model3} 
\end{eqnarray}
where the time variable has been normalized to the photon lifetime and $x_i, y_i$ are respectively the dimensionless photon and carrier density of the laser $i$. $X = \frac{1}{N}\sum\limits_{i=1}^N x_i$ is the total intensity normalized to the number of elements in the network. The global variable $w$ is the (scaled) high-pass filtered feedback current, which includes a saturable nonlinear function $f(X) = A\ln(1+\alpha X)$. 

\section{Theory}
Throughout this section, $I$ denotes a subset of the integers $[ 1,N ]$ of cardinal $N_+$. It will be used to label the lasers which are switched ON.
\subsection{Stationary states}
We start by computing the stationary solutions of \eqref{eq:model1} when $N$ lasers are connected. Let us write $P_I$  the equilibrium for which $\forall i\in I, x_i>0$ and $x_i=0$ otherwise. One also define $X_I^{eq} =\frac1N\sum\limits_{i\in I}x_i$. We first have $0=w+f(X_I^{eq})$.

\begin{itemize}
	\item  Case $i\in I$. One finds that $x_i = \delta_i-1$ and $y_i=1$.  
	\item  Case $i\notin I$. It gives $x_i=0$ and $y_i = \delta_i$
\end{itemize} 
Finally $w=-f(\frac1N\sum\limits_{i\in I}\delta_i-\frac{N_+}{N})$. Note that there are $2^N$ such equilibria.

\subsubsection{Stability of the stationary states}
In this section, we compute the stability of the equilibria $P_I$. To this end, we write $x_i(t) = x_i^{eq} + e^{\lambda t}u_i$,  $y_i(t) = y_i^{eq} + e^{\lambda t}v_i$ and $w(t) = w^{eq} + e^{\lambda t}\omega$. We then Taylor expand \eqref{eq:model1}-\eqref{eq:model3} under the assumption that $u_i,v_i,\omega$ are small. The goal is to find the eigenvalues $\lambda$ associated to a non trivial set of $u_i,v_i,\omega$. One gets $X_I^{eq}:=\frac1N\sum_ix_i^{eq}$ and
\begin{eqnarray*}
	\lambda u_i & = & x_i^{eq}v_i+u_i(y_i^{eq}-1) \\\
	\lambda v_i & = & \gamma(-v_i-x_i^{eq}v_i-u_iy_i^{eq}+k(\omega+f'(X_I^{eq}) \frac1N\sum\limits_ku_k) )\\
	\lambda \omega  & = & -\epsilon(\omega+f'(X_I^{eq}) \frac1N\sum\limits_ku_k).
\end{eqnarray*}
which gives
\begin{eqnarray}\label{eq:eq-vp}
	(	\lambda-y_i^{eq}+1) u_i & = & x_i^{eq}v_i \label{eq:eq-vp1}\\
	(\lambda+\gamma+\gamma x_i^{eq}) v_i & = & -\gamma u_iy_i^{eq}-\gamma k\frac\lambda\epsilon \omega \label{eq:eq-vp2}\\
	(\lambda+\epsilon) \omega  & = & -\epsilon f'(X^{eq}_I) \frac1N\sum\limits_ku_k\label{eq:eq-vp3}.
\end{eqnarray}

We can find some eigenvalues analytically. Indeed, if we consider $i_0\notin I$, then $x^{eq}_{i_0}=0$. This gives the following cases:
\begin{itemize}
	\item Case $u_{i_0}=1$. We find that $\lambda = y^{eq}_{i_0}-1$ is an eigenvalue. Indeed, set $u_i=v_i=0$ for $i\neq i_0$. Then $\omega = - \frac1N\frac{\epsilon}{\lambda+\epsilon}f'(X^{eq})$ and $v_{i_0}$ is found using equation \eqref{eq:eq-vp2}. This gives $N-N_+$ eigenvalues.
	\item Case $u_{i_0}=0$. We find that $\lambda = -\gamma$ is an eigenvalue. Indeed, set $u_i=v_i=0$ for $i\neq i_0$. Then $\omega=0$ and $v_{i_0}=1$ is solution of the above equations. This gives $N-N_+$ eigenvalues.
\end{itemize}
We now look for the remaining eigenvalues. Using the above equations \eqref{eq:eq-vp1}-\eqref{eq:eq-vp3}, one finds:
\[ \left(\lambda-y_i^{eq}+1+\frac{\gamma x_i^{eq}y_i^{eq}}{\lambda+\gamma+\gamma x_i^{eq}} \right) u_i = \frac{x_i^{eq}}{\lambda+\gamma+\gamma x_i^{eq}} \frac{\lambda}{\lambda+\epsilon}\left[ \frac{\tilde A_I}{N}\sum_ku_k \right]  \]
where we wrote $\tilde A_I := \gamma kf'(X^{eq}_I)$. 
By summing this equation w.r.t. $i$, one extracts an equation for $\sum_i u_i\neq 0$ and get
\[ 1=\frac{\tilde A_I}{N}\frac{\lambda}{\lambda+\epsilon} \sum\limits_{i=1}^N \frac{x_i^{eq}}{\gamma x_i^{eq}y_i^{eq}+(\lambda+\gamma+\gamma x_i^{eq})(\lambda-y_i^{eq}+1)}. \]
We can simplify this equation in the case of the equilibrium $P_I$ when $\lambda+\gamma\neq0$ and $\lambda\neq\delta_i-1$ for $i\notin I$ into an equation accounting for the dynamics of the switched ON lasers:
\begin{equation} 
\label{eq:VP}
1=\frac{\tilde A_I}{N}\frac{\lambda}{\lambda+\epsilon} \sum\limits_{i\in I} \frac{x_i^{eq}}{\gamma x_i^{eq}+\lambda(\lambda+\gamma+\gamma x_i^{eq})}. 
\end{equation}
Note that this polynomial equation has at most $2N_++1$ zeros which in addition to the other $2(N-N_+)$ zeros gives $2N+1$ eigenvalues as expected.

\subsubsection{Approximation of small deviation}

Solving the previous equation \eqref{eq:VP} is tedious but we can simplify it.
Let us assume that $\delta_i = \Delta+z\eta_i$ where $z<<1$, \textit{i.e.} the control parameters are peaked around $\Delta$. We use the fact that $x^{eq}_i = \Delta-1+z\eta_i$ for $i\in I$ and rewrite \eqref{eq:VP} as $1=P(\lambda,z)$. Also, we write $\tilde A_I = \tilde A_I^0+z\tilde A_I^1=\gamma kf'(\frac{N_+}{N}(\Delta-1))+z\gamma kf''(\frac{N_+}{N}(\Delta-1))\frac1N\sum\limits_{i\in I}\eta_i$.

Using Maple, we Taylor expand $P(\lambda,z)$ in $z$ at first order:
\begin{equation*} 
1=\frac{1}{N}\frac{\lambda}{\lambda+\epsilon} \left( (\tilde A_I^0+z\tilde A_I^1)N_+\frac{\Delta -1}{\gamma (\Delta-1)+\lambda(\lambda+\gamma+\gamma (\Delta-1))}
 +z\frac{\tilde A_I^0\lambda(\lambda+\gamma)}{\left[\gamma (\Delta-1)+\lambda(\lambda+\gamma+\gamma (\Delta-1))\right]^2} \sum\limits_{i\in I}\eta_i\right). 
\end{equation*}
We solve this equation perturbatively by seeking $\lambda = \lambda_0 + z\lambda_1+O(z^2)$. One gets 
\[ 1=  \tilde A_I^0\frac{N_+}{N}\frac{\lambda_0}{\lambda_0+\epsilon}\frac{\Delta -1}{\gamma (\Delta -1)+\lambda_0(\lambda_0+\gamma+\gamma (\Delta -1))} \]
and
\[
\lambda_1 \propto\sum\limits_{i\in I} \eta_i.
\]
The first equation in $\lambda_0$ is solved similarly to the single laser case. When $N_+=N$ (all lasers are ON), the $x_i^{eq}$ solve the same equations as for the isolated laser with common control parameter $\Delta$. 

{If we chose $\Delta = \frac1N\sum_{i=1}^N\delta_i$, then one finds that $ \frac1N\sum_{i=1}^N\eta_i=0$ and thus $\lambda_0$ is precise at second order in $z$. The second order correction to $\lambda$ is then function of the second moment $\sum_{i\in I}\eta_i^2$.}

\subsection{Critical Manifold $S_I$}
The critical manifold is defined by solving for each $w$ the following equations:
\begin{eqnarray}
0 & = & x_i(y_i-1) \\\
0 & = & \gamma(\delta_i-y_i+k(w+f(X)) - x_iy_i)
\label{eq:criticalmanifold}
\end{eqnarray}

As before, we parameterize the critical manifold by the set $I$ which labels the switched ON lasers. We denote by $S_I$ the associated critical manifold. Note that the critical manifold is composed of $2^N$ components, namely: $S=\cup_{I\subset[1,N]}S_I$. Using the same arguments as for the equilibria, it is straightforward to show that 
\[S_I = \{ (x^I_i(w),y^I_i(w),w),\ i=1\cdots N,\ w\in\mathbb R\}\]
with
\[
(x_i^I(w),y^I_i(w)) = \left\lbrace  
\begin{aligned}
(0,\delta_i+k(w+f(X_I(w)))),&\quad \forall i\notin I,\\
(\delta_i-1+k(w+f(X_I(w))),1)&\quad \forall i\in I,
 \end{aligned}\right.
  \]
where $X_I(w)$ is implicitly defined by
\begin{equation}
\label{eq:XI}
X_I(w) = \frac{N_+}{N}(k(w+f(X_I(w)))-1)+\frac1N\sum\limits_{i\in I}\delta_i. 
\end{equation}

\subsubsection{Stability of the critical manifold}
In this section, we compute the eigenvalues of the linearized equation around the critical manifold when $\epsilon=0$:
\begin{eqnarray}
\dot x_i & = & x_i(y_i-1) \\\
\dot y_i & = & \gamma(\delta_i-y_i+k(w+f(X)) - x_iy_i).
\end{eqnarray}
To this end, we write $x_i(t) = x_i^I(w) + e^{\lambda t}u_i$ and  $y_i(t) = y_i^I(w) + e^{\lambda t}v_i$ and Taylor expand the above equation with the assumption that $u_i,v_i$ are small. The goal is to find $\lambda$ associated to a non trivial set of $u_i,v_i$. One gets:
\begin{eqnarray*}
	\lambda u_i & = & x_i^Iv_i+u_i(y_i^I-1) \\
	\lambda v_i & = & \gamma(-v_i-x_i^Iv_i-u_iy_i^I+kf'(X_I(w)) \frac1N\sum\limits_ku_k ).
\end{eqnarray*}
which gives
\begin{eqnarray}
	(	\lambda-y_i^I+1) u_i & = & x_i^Iv_i \label{eq:crit-vp1}\\
	(\lambda+\gamma+\gamma x_i^I) v_i & = & -\gamma u_iy_i^I+\gamma kf'(X_I(w)) \frac1N\sum\limits_ku_k.\label{eq:crit-vp2}
\end{eqnarray}

We can find some eigenvalues analytically. Indeed, if we consider $i_0\notin I$, then $x^{I}_{i_0}(w)=0$. This gives the following cases:
\begin{itemize}
	\item Case $u_{i_0}=1$. We find that $\lambda = y^{I}_{i_0}(w)-1$ is an eigenvalue. Indeed, set $u_i=v_i=0$ for $i\neq i_0$. Then $v_{i_0}$ is found using equation \eqref{eq:crit-vp1}. This gives $N-N_+$ eigenvalues.
	\item Case $u_{i_0}=0$. We find that $\lambda = -\gamma$ is an eigenvalue. Indeed, set $u_i=v_i=0$ for $i\neq i_0$. Then $v_{i_0}=1$ is solution of the above equation. This gives $N-N_+$ eigenvalues.
\end{itemize}
We now look for the remaining eigenvalues. Using the above equations \eqref{eq:crit-vp1}-\eqref{eq:crit-vp2}, one finds:
\[ \left(\lambda-y_i^I+1+\frac{\gamma x_i^Iy_i^I}{\lambda+\gamma+\gamma x_i^I} \right) u_i = \frac{x_i^I}{\lambda+\gamma+\gamma x_i^I} \left[ \frac{\tilde A_I(w)}{N}\sum_ku_k \right]  \]
where we wrote $\tilde A_I(w) := \gamma kf'(X_I(w))$. 
By summing this previous equation w.r.t. $i$, one extracts an equation for $\sum_i u_i\neq 0$:
\[ 1=\frac{\tilde A_I(w)}{N} \sum\limits_{i=1}^N \frac{x_i^I(w)}{\gamma x_i^I(w)y_i^I(w)+(\lambda+\gamma+\gamma x_i^I(w))(\lambda-y_i^I(w)+1)}. \]
We can simplify this equation because $(x_i^I,y_i^I)$ belongs to $S_I$ and when $\lambda+\gamma\neq0$, $\lambda\neq y_i^I(w)-1$ (for $i\notin I$):
\begin{equation} 
1=\frac{\tilde A_I(w)}{N} \sum\limits_{i\in I} \frac{x_i(w)}{\gamma x_i(w)+\lambda(\lambda+\gamma+\gamma x_i(w))}. 
\label{eq:stabcriticalmanifold}
\end{equation}
This provides an equation for the remaining $2N_+$ eigenvalues.

\subsubsection{Small deviation approximation of the critical manifold}
For notation purposes, we write $x_i^I(w) = x_i(w)$ and $y_i^I(w) = y_i(w)$.

\

Solving the previous equation \eqref{eq:stabcriticalmanifold} is tedious but we can simplify it. Let us assume that $\delta_i = \Delta+z\eta_i$ where $z<<1$, \textit{i.e.} the current values are peaked around $\Delta$. Our goal is to Taylor expand  \eqref{eq:stabcriticalmanifold} in $z$ and solve it perturbatively. Hence, we need to find $x^I_i(w),y^I_i(w)$ as function of $z$.

We write $\forall i\in I,\ x_i(w) = x_i^0(w) + zx_i^1(w)+O(z^2)$ and note that $y_i(w)=1$. To find these expressions, we need to find the following expressions $X_I(w)=X_I^0(w)+zX_I^1(w)+O(z^2)$. We would then have $x_i^0(w) = \Delta-1+k(w+f(X_I^0(w)))$ which is independent of $i$, and so is written $x^0_I(w)$, and $x_i^1(w) = \eta_i+kf'(X_I^0)X_I^1$. Using \eqref{eq:XI}, we find that $X_I^0(w),X_I^1(w)$ solves:
\[
X_I^0(w) = \frac{N_+}{N}(k(w+f(X_I^0(w)))-1)+\Delta
\]
and
\[
X_I^1(w) = \frac{N_+}{N}kf'(X_I^0(w))X_I^1+\frac1N\sum_I\eta_i\ \Rightarrow X^1_I(w) = \frac{\frac1N\sum_I\eta_i}{1-\frac{N_+}{N}kf'(X_I^0(w))}.
\]
We obtain the following expression
\[
x_i^1(w) = \eta_i+\frac{kf'(X_I^0(w))}{1-\frac{N_+}{N}kf'(X_I^0(w))}\frac1N\sum_I\eta_i.
\]
Note that the equation for $x_I^0$ is exactly the same equation as for the single laser \textit{i.e.} but with parameters $\alpha,\delta$ changed into $\alpha\frac{N_+}{N}$ and $\Delta$.
\subsubsection{Small deviation approximation of the stability of critical manifold}
We now proceed to find the stability of the critical manifold using \eqref{eq:stabcriticalmanifold}. As before, we write $\tilde A_I(w) = \tilde A_I^0(w)+z\tilde A_I^1(w)=\gamma kf'(X_I^0(w))+z\gamma kf''(X_I^0(w))X_I^1(w)$.
We Taylor expand \eqref{eq:stabcriticalmanifold} in $z$ and find:
\[
1 = (\tilde A_I^0(w)+z\tilde A_I^1(w)) \frac{N_+}{N}\frac{x^0_I(w)}{\gamma x^0_I(w)+\lambda(\lambda+\gamma+\gamma x^0_I(w))} 
+z\frac{\lambda(\lambda+\gamma)}{\left[\gamma x^0_I(w)+\lambda(\lambda+\gamma+\gamma x^0_I(w))\right]^2} \frac{\tilde A_I^0(w)}{N}\sum\limits_{i\in I} x_i^1(w). 
\]
We solve this equation perturbatively by seeking $\lambda = \lambda_0 + z\lambda_1+o(z)$. One gets 
\begin{equation}\label{eq:crit-stab0th}
 1=  \tilde A_I^0(w) \frac{N_+}{N}\frac{x^0_I(w)}{\gamma x^0_I(w)+\lambda_0(\lambda_0+\gamma+\gamma x^0_I(w))} 
 \end{equation}
and
\[\lambda_1 \propto\sum\limits_{i\in I} \eta_i.
\]
The first equation in $\lambda_0$ is quadratic and easily solved. In the case $N_+=N$, as $x_I^0(w)$ solves the same equations as for the isolated laser, one finds that $\lambda_0$ solves the same equation for the stability of the critical manifold. In effect, those two terms corresponds to the single laser case. 

\

At zero order in $z$, the components $(x_i(w),y_i(w))$ for $i\in I$ of the critical manifold $S_I$ are all the same and share the expression of the critical manifold (with $x>0$) of the single laser with control parameter $\Delta$ and $\alpha\to\alpha\frac{N_+}{N}$. 
When $N_+=N$ and at zero order in $z$, the stability of the $S_{[1,N]}$ branch is therefore the same as that of an uncoupled laser with control parameter $\Delta$. 
However, when $N_+<N$, the other eigenvalues $y_i(w)-1$ may influence the stability of $S_I$. 

\subsubsection{Summary}
The above discussion hints at introducing the laser dynamics
\begin{eqnarray}
\dot{x} & = & x(y-1) \label{eq:MF0th1}\\
\dot{y} & = & \gamma\left(\Delta-y+k\left(w+f\left(\frac{N_+}{N}x\right)\right) - xy\right)\\
\dot{w} & = & -\epsilon\left(w + f\left(\frac{N_+}{N}x\right)\right) \label{eq:MF0th3}
\end{eqnarray}

with critical manifold composed of the OFF branch $(0,\Delta+kw)$ and the ON branch $(x^{crit},1)$ where $x^{crit}$ solves \[x^{crit}(w) = \Delta-1+k\left( w+f\left(\frac{N_+}{N}x^{crit}(w)\right)\right).\]
The eigenvalues along the OFF branch are $\Delta+kw-1$ and $-\gamma$.
The eigenvalues $\lambda$ along the ON branch are solutions of 
\begin{equation}\label{eq:stab-MFlaser}
\lambda(\lambda+\gamma+\gamma x^{crit}(w))-\gamma\left(k\frac{N_+}{N}f'\left(\frac{N_+}{N}x^{crit}(w)\right)-1\right)x^{crit}=0
\end{equation}

At zeroth order in $z$, the critical manifold $S_I$ is given by $(x_i^I(w),y^I_i(w)) = \left(0, \Delta+kw+\textcolor{red}{kf\left(\frac{N_+}{N}x^{crit}(w)\right)}\right)$ for $i\notin I$ and $(x_i^I(w),y^I_i(w)) = (x^{crit}(w),1)$ for $i\in I$. Hence, only the OFF part of the critical manifold $S_I$ differs from the above model with the correction shown in red. For the stability of $S_I$, we have the eigenvalues $-\gamma,\Delta+kw+\textcolor{red}{kf\left(\frac{N_+}{N}x^{crit}(w)\right)}-1$ and the solutions of \eqref{eq:stab-MFlaser} (which are the same as those of \eqref{eq:crit-stab0th}). Hence, only the OFF part of the dynamics is inadequately described by \eqref{eq:MF0th1}-\eqref{eq:MF0th3}.